\documentclass[runningheads]{llncs}
\usepackage{graphicx}
\usepackage{amsmath,amssymb} 
\usepackage{color}
\usepackage[normalem]{ulem}
\useunder{\uline}{\ul}{}

\usepackage[linesnumbered,ruled,vlined]{algorithm2e}

\usepackage{graphicx}
\usepackage{amsmath}
\usepackage{float}
\usepackage{hyperref}

\usepackage{multirow}
\usepackage[table,xcdraw]{xcolor}

\usepackage{hyperref}
\hypersetup{
	colorlinks=true,  
	linkcolor=black,  
	citecolor=black,  
	urlcolor=black    
}

\DeclareMathOperator*{\argmax}{arg\,max}
\DeclareMathOperator*{\argmin}{arg\,min}
\usepackage[width=122mm,left=12mm,paperwidth=146mm,height=193mm,top=12mm,paperheight=217mm]{geometry}
\begin{document}
	\pagestyle{headings}
	\mainmatter
	\def\ECCV16SubNumber{4692}  
	
	\title{Counter-Samples: A Stateless Strategy to Neutralize Black Box Adversarial Attacks} 
	

	\titlerunning{ECCV-16 submission ID \ECCV16SubNumbern}
	
	\authorrunning{ECCV-16 submission ID \ECCV16SubNumber}
	
	\author{Roey Bokobza
        \and
        Yisroel Mirsky}
	\institute{Ben-Gurion University \\
Department of Software and Information Systems Engineering\\Offensive AI Research Lab}

	\maketitle
	
	\begin{abstract}
		Our paper presents a novel defence against black box attacks, where attackers use the victim model as an oracle to craft their adversarial examples. Unlike traditional preprocessing defences that rely on sanitizing input samples, our stateless strategy counters the attack process itself. For every query we evaluate a counter-sample instead, where the counter-sample is the original sample optimized against the attacker's objective. By countering every black box query with a targeted white box optimization, our strategy effectively introduces an asymmetry to the game to the defender's advantage. This defence not only effectively misleads the attacker's search for an adversarial example, it also preserves the model's accuracy on legitimate inputs and is generic to multiple types of attacks. 
		
		We demonstrate that our approach is remarkably effective against state-of-the-art black box attacks and outperforms existing defences for both the CIFAR-10 and ImageNet datasets. Additionally, we also show that the proposed defence is robust against strong adversaries as well.
	\end{abstract}

	\section{Introduction}
	Deep neural networks are susceptible to adversarial examples \cite{szegedy2013intriguing,goodfellow2014explaining}. Adversarial examples are inputs to machine learning models that have been intentionally perturbed in a subtle manner to induce a misclassification. These perturbations, denoted by \( \delta \), are achieved by solving the following optimization objective 
	\begin{equation}\label{eq:adv_obj}
		\delta^* = \argmin_{\delta} \| \delta \|_p \quad \text{subject to} \quad f(x + \delta) \neq f(x) \quad \text{and} \quad \| \delta \|_p \leq \epsilon
	\end{equation}
	Here, \(f\) represents the model under attack, and the condition \( \|\delta \|_p \leq \epsilon \) ensures that the adversarial example \(x'=x+\delta\) is visually indistinguishable from \(x\). This objective holds whether the attacker is using a search-based algorithm or gradient-based optimization algorithm to find $\delta$. 
	
	Numerous algorithms for finding adversarial examples have been proposed over the years. In a white box setting, the attacker has full access to the model's parameters, enabling the attacker to compute gradients directly over the parameters of $f$ to find $\delta$. This approach is used in popular attacks such as PGD \cite{PGD}, FGSM \cite{goodfellow2014explaining}, and C\&W \cite{carlini2017towards}. However, in numerous real-world applications, the scenarios are predominantly black box in nature, meaning that the attacker does not have access to the model's internal parameters. This is especially common in situations where the model is deployed over the Internet or integrated within a product, limiting the attacker's interaction with the model to only querying it. Instead, the adversary must resort to other tactics to find $\delta$. 
	
	A common black box approach is to use a posteriori information; by querying the victim, is possible to measure the loss at $f(x)$ \cite{ZosignSGD} and estimate the optimal direction for updating $\delta$. By repeating this process, an attacker can incrementally optimize $\delta$ to satisfy the objective. These attacks are referred to as query-based black box attacks.
	
	We observed that in each iteration of a query-based attack, the algorithm consistently aims to identify the direction towards the nearest boundary. Guided by this insight, we have the following question: \textit{Is it possible to mislead the attacker by evaluating \(f\) at a location entirely different from what the attacker intended?} Doing so would render the search directions estimated by the adversary at each step arbitrary, resulting in a failed attack due to an excessive number of queries.
	
\begin{figure}[t]
	\centering
	\includegraphics[width=\textwidth]{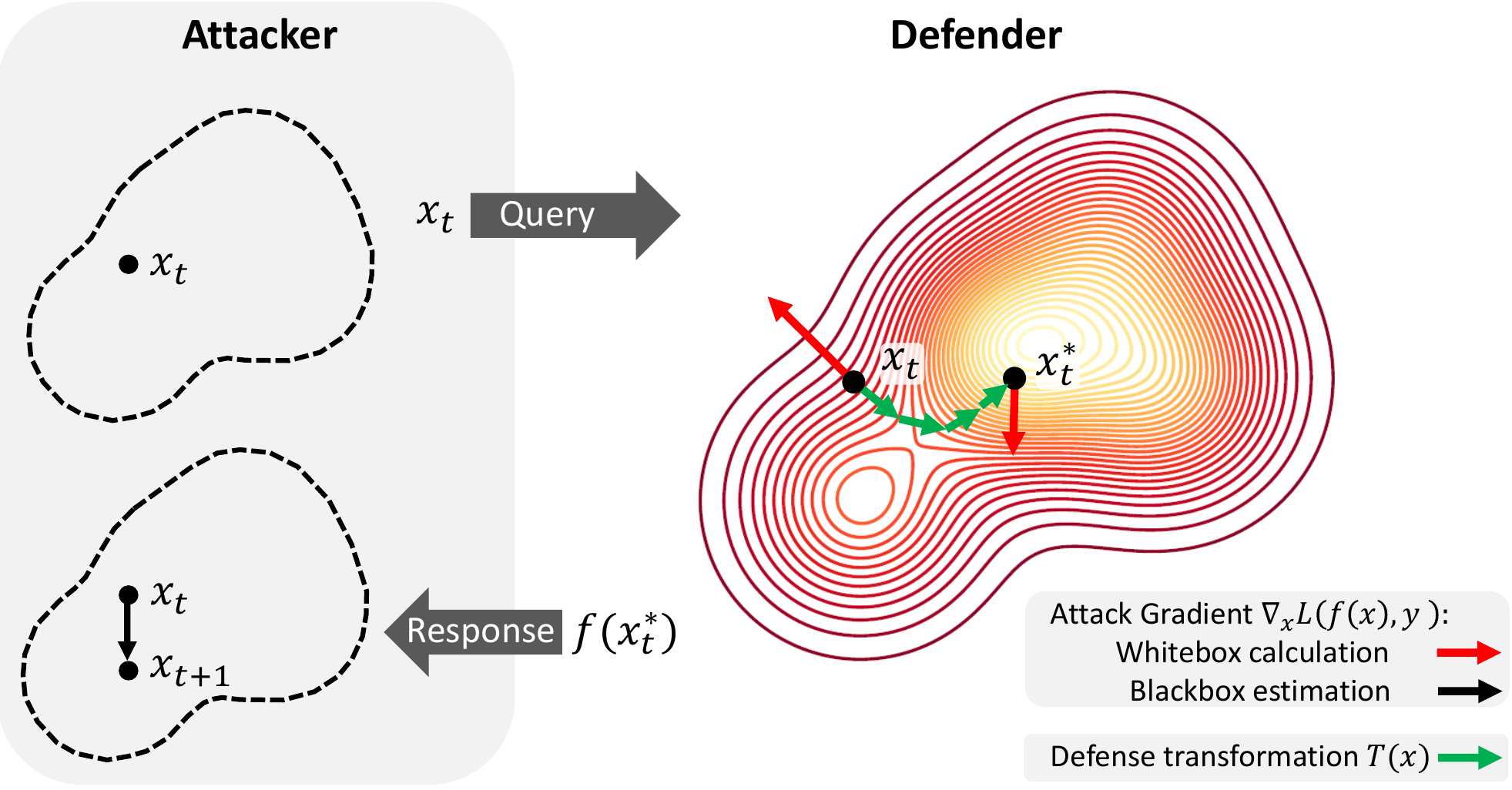}
	\caption{An illustration of how the counter-sample defence works. For each attack iteration, (1) the adversary sends one sample $x_t$ to estimate the gradient or direction that will maximizes the loss, (2) the defender applies the preprocessor $T(x_t)=x_{t}^{*}$ which uses gradient descent to distance $x_t$ away from the attacker's objective and (3) the attacker receives response $f(x_{t}^{*})$ and estimates the wrong direction for $x_{t+1}$ as a result.}
	\label{fig:intro}
\end{figure}

	In this paper, we propose a novel preprocessor which accomplishes this goal (illustrated in Fig. \ref{fig:intro}): given an input $x$ we generate a counter-sample $x^*$ by simply making $x$ appear more like its perceived class. This can be achieved with very few iterations of gradient descent.
	If the attacker's initial query does not result in an adversarial sample, then this approach can effectively confine the search within the class's manifold.  

	Our approach to creating-counter examples introduces several key advantages:
	
	\begin{description}
		\item[Preservation of Clean Task Performance:] Unlike many existing defence \\mechanisms that degrade the model's performance on legitimate inputs \\through methods like additive noise, blurring, or compression \cite{RND,GaussianNoiseSND,JpegCompression,guo2017countering,FeatureSqueezing}, our strategy maintains the integrity of the model's predictions on benign samples. Our pre-processing step ensures that the input is modified in a way that does not alter the present prediction, thus preserving the model's accuracy.
		\item[Asymmetry in Optimization Capabilities:] Our defence exploits the discrepancy between the attacker's and the defender's capabilities. In a black box scenario, the attacker is limited to making iterative queries to approximate the direction that minimizes the loss for (\ref{eq:adv_obj}) \cite{BackinBlack}. In contrast, the defender can perform multiple white-box optimization steps on the input before it is processed by the model. This not only puts the attacker at a disadvantage but also severely limits the efficacy of attacks that rely on multiple queries for a single optimization step, such as the ZOO attack \cite{ZosignSGD}. Moreover, while the adversary is constrained to providing samples that adhere to feature bounds (e.g., legal pixel ranges), the defender faces no such limitations when crafting a counter-sample. Additionally, by taking a few gradient descent steps in a direction contrary to the attacker's objective, the gradient's magnitude can be reduced. This subtle manipulation makes can make it more challenging for the attacker to navigate out of the misled region.
		\item[Statelessness and Scalability:] The defence does not require tracking the history of a user's queries to detect malicious patterns like is done in these works \cite{chen2020stateful,meng2017detector}. This feature greatly enhances the scalability of counter-samples. This statelessness ensures that the defence can be seamlessly integrated into large-scale systems without the overhead of maintaining state or history for each interaction. It simplifies the defence mechanism while making it broadly applicable and easier to manage in practice.
	\end{description}
	
	In summary, our paper introduces a novel stateless defence against query-based black box adversarial attacks. The proposed strategy, preserves model accuracy on legitimate inputs while exploiting optimization asymmetries between attackers and defenders. It offers a scalable defence mechanism, and marks a significant advancement in securing AI systems against adversarial threats. Finally, to the best of our knowledge, this is the first work that proposes a stateless defence against query-based black box attacks that directly counters the adversarial objective.

We evaluate our method against 10 query-based black box attacks over the CIFAR-10 and ImageNet datasets. We also compared our performance to 7 defences. The proposed defence achieves state-of-the-art results while effective in preserving the clean task performance, unlike other defences. We also analyze our algorithm's hyperparameters and evaluate the defence against adaptive adversaries to demonstrate the method's robustness.

\section{Related Work}
Defending machine learning models against adversarial examples is a critical area of research, with strategies broadly falling into three categories: preprocessors, detectors, and model hardening techniques. Preprocessors offer several advantages; unlike detectors, they do not require blocking samples, making them less disruptive. They are also more efficient and practical than model hardening approaches, such as adversarial retraining, and generalize well across different models without the need for extensive retraining or modification \cite{ReactiveProactive}.

The defence proposed in this paper is a preprocessors which is designed to counter query-based black box attacks. Therefore, we will first review prerpocessors in general, and then focus on black box specific defences.

\textbf{Preprocessors.} A preprocessor modifies a sample before it is executed by a model in order to mitigate the impact of any potential perturbations. Some preprocessors aim to remove perturbations by either compressing the feature space \cite{JpegCompression}, removing high frequency features \cite{FeatureSqueezing}  or by injecting additive noise \cite{RND,GaussianNoiseSND}. Others try to avoid pertubations by replacing samples with a similar ones \cite{PixelDefend,MagNET,DefenseGAN}. This technique often involves applying an auxiliary model to every sample before it is executed. 

Existing preprocessors have two major disadvantages: (1) they are known to be weak against adaptive adversaries which consider the processor during the attack \cite{FalseSecurity} and (2) they harm the model's clean task performance. In contrast, our method is robust against adaptive adversaries because it dynamically responds to each black box query throughout the attack generation algorithm, and not just the final adversarial example $x'$. Moreover, our method has a negligible affect on clean task performance making the defence highly desirable in a production setting.

\textbf{Black box Specific Defences.}
Black box attacks operate by leveraging either a priori knowledge, such as transferability via substitute models, or a posteriori knowledge by  querying the victim model as an oracle to iteratively refine $\delta$  \cite{BackinBlack}. It is critical to underscore that a priori and a posteriori methods are orthogonal to each other, allowing for their combination in more sophisticated attack strategies (e.g., \cite{HOGA}). In this work, we focus on a posteriori attacks, a.k.a query-based attacks.

Defences against query-based attacks can be categorized as being either stateful or stateless. For example, in \cite{chen2020stateful} the authors showed how it's possible to detect when a user is trying to create an adversarial example by monitoring the user's query history; as the algorithm progresses, each subsequent query becomes increasingly adversarial. Unfortunately, stateful defences do not scale well to large systems and assume that attackers will perform the entire attack from a single user account. Researchers have investigated stateless defences as well. 
In \cite{RND,GaussianNoiseSND} the authors investigate the impact of additive noise as a preprocessors against query-based attacks. Through a comprehensive theoretical analysis, the work shows how the inference disrupts the attacker's gradient estimation and slows the adversarial example generation process. However, random noise can be mitigated by an adaptive adversary who simply averages repeated queries. Furthermore, additive noise is a static defence which enables the attacker to adapt its strategy which optimizing $x'$. In contrast, our method is robust to adaptive adversaries because it dynamically responds to each query while remaining stateless; having no dependence on prior queries.

\section{Methodology}\label{sec:meth}

\subsection{Problem Statement}
Our objective is to alter an input sample \(x\) into a transformed variant \(x^*\) called a counter-sample. The transformation should ensure that \(x^*\) upholds the hard label prediction of \(x\) as determined by the classification model \(f\), while also presenting a markedly different loss surface in its vicinity. This strategy is designed to complicate the black box process attackers use to craft adversarial examples, thus reducing their efficacy.

Given the unavailability of the true label \(y\) to the defender, we depend on the model \(f\) to predict the label for \(x\), denoted by \(\hat{y} = \argmax_i f(x)_i\). We operate under the assumption that the initial query in an attack scenario will not be adversarial, a premise supported by the nature of query-based black box attacks which necessitate numerous queries to sufficiently alter the original sample \(x_0\) into an adversarial one \(x'\) \cite{BackinBlack}.

With this premise, our problem is defined as follows: Given a classification model \(f\) with parameters \(\theta\), an input sample \(x\), and a loss function \(L(f(x;\theta), \hat{y})\) utilized for training \(f\), our goal is to generate a transformed sample \(x^*\) where \(\argmax_i f(x)_i = \argmax_i f(x^*)_i\) and the local loss surface of \(f(x^*)\) will significantly diverges from that of \(f(x)\).

\subsection{Method}
We propose one possible implementation; a gradient optimization strategy that iteratively adjusts \(x\) towards the label $\hat{y}$. The gradient descent formulation for our defence mechanism is as follows:

\begin{equation}\label{eq:defence}
	x^{*}_{t+1} = x^{*}_{t} - \alpha \nabla_{x^{*}_t} \mathcal{L}(f(x^{*}_t;\theta), \hat{y})
\end{equation}

In this expression, \(x^{*}_{t}\) signifies the approximation of \(x^*\) at iteration \(t\), \(\alpha\) is the learning rate, and \(\nabla_{x^{*}_t} \mathcal{L}\) represents the gradient of the loss function \(\mathcal{L}\) with respect to \(x^{*}_t\). The loss function \(\mathcal{L}\) is assessed against the model's prediction \(f(x^{*}_t;\theta)\) and the hard label prediction \(\hat{y}\), directing \(x^{*}\) to more closely align with its predicted class and away from potential adversarial susceptibilities. 

This process of gradient optimization continues until \(x^{*}_t\) satisfactorily converges to \(x^*\), determined by either reaching a maximum number of iterations or achieving a minimal threshold in the variation of \(x^{*}_t\) across iterations. The result is a counter sample $x^*$ that not only has a different loss surface than $x$, but also has a much smaller gradient.
This approach maintains the model's performance on legitimate inputs but also substantially diminishes the adversary's ability to explore the loss surface at each attack iteration.

Although our method is gradient-based, it affects search based attacks as well (e.g. Simba \cite{SimBA}). This is because (\ref{eq:defence}) gives $x$ a new loss surface which misinforms the search algorithm.

\subsection{Positive Sum Game} \label{SumGame}
\textbf{Formal Definition of the Game.}
In the adversarial context we described, the game unfolds with the attacker and defender taking turns to query the model. The attacker is permitted a single black box query to the model to decide on the adjustment of \(x\), aiming to craft an adversarial example. The defender is allowed unlimited white box queries with \(x\) to generate a response designed to mislead the attacker. This iterative process continues, where each party seeks to outmaneuver the other within the constraints of their capabilities.

\noindent\textbf{Asymmetry Advantage.}
Contrary to what might initially appear as a zero-sum game—where gains and losses between adversaries are perfectly balanced—the scenario we outline is inherently asymmetric, benefiting the defender. This asymmetry arises because the defender can engage in multiple iterations of adjustments for every single attack attempt, effectively employing our algorithm to counter each query from the attacker. It is crucial to note that an attacker's query might not directly constitute an attack iteration but rather a component of it, possibly just one step in computing a gradient direction.

In this setting, even as the attacker endeavours to step towards a decision boundary, the defender has the capability to take several steps \textit{back} for every forward step the attacker takes, thereby aiming to distance the attacker from meaningful gradient information. This advantage, however, is not absolute, as the possibility exists for an attacker to cross the boundary on their initial attempt, depending on the proximity of the sample to the class boundary and the effectiveness of the attack strategy.

Moreover, while the attacker is constrained to adjust their perturbations within legal pixel value ranges, the defender has no such limitation. This provides the defender with a broader spectrum of defensive maneuvers. This disparity grants the defender a significant strategic advantage, providing a wider array of options to counteract adversarial attempts.

\subsection{Noise Extension}\label{subsec:noise}
Because of $\hat{y}$, the approach in (\ref{eq:defence}) becomes ineffective the moment the adversary manages to cross the boundary. However, it has been shown that by adding uniform noise to a weak adversarial example, it is possible to push it back over the boundary \cite{hedge}. Therefore, we strengthen our approach further by adding uniform noise to each query before executing (\ref{eq:defence}). 

Moreover, by including additive noise, we also extend our defence to black box decision-based attacks as well, such as \cite{HopSkipJump,GeoD,SignOPT}. Although these attack do not rely on exploring loss surfaces, they are highly susceptible to noise \cite{RND,GaussianNoiseSND}. Thus, our complete defence of the model $f$ can be expressed as
\begin{equation}\label{eq:full_defence}
	f'(x) = f(T(x+z);\theta)
\end{equation}
where $T$ is the transformation algorithm in (\ref{eq:defence}) and $z$ is a vector of random Guassian noise. The complete defence algorithm is outlined in Algorithm \ref{algo:cs}.

\begin{algorithm}[t]
\DontPrintSemicolon 
\KwIn{$x$ (received query), $k$ (number of iterations), $\alpha$ (step size), \\$\mathcal{L}$ (Loss function), $f$ (model)}
\KwOut{$x^*$ (counter-sample)}
$z \sim \mathcal{N}(\mu, \sigma^2)$\;
$x^*_{1} \gets x + z$\;
\For{$i \gets 1$ \textbf{to} $k$} {
  $\hat{y} = \argmax_j f(x^*)_j$ \;
   $x^*_{i+1} \gets x^*_{i} - \alpha \times \nabla_{x^*_{i}} \mathcal{L}( f(x^*_{i}),\hat{y})$ \;
  }
\Return{$x^*_{k}$}\;
\caption{Counter-Sample Creation}
\label{algo:cs}
\end{algorithm}


\section{Experiment Setup}
\textbf{Datasets and Models.} To evaluate our defence, we followed the same approach as done by Zeyu Qin et al. in \cite{RND}. We used 10,000 random samples from the test set of CIFAR-10 \cite{CIFAR10} and 1,000 random samples from the test set of ImageNet \cite{ImageNet} for generating the adversarial examples and evaluating the performance of our method. For CIFAR-10, we used a pretrained ResNet-20 model taken from PyTorch hub library,\footnote{\url{https://pytorch.org/hub}} and for ImageNet, we used a pretrained ResNet-50 model from TorchVision library.\footnote{\url{https://github.com/pytorch/vision}}


\noindent\textbf{Attack Algorithms.} 
We evaluated a variety of 10 state-of-the-art query-based black box attacks on our defence. For attacks that estimate the gradient, we used ZOSignSGD (ZS) \cite{ZosignSGD}, Parsiminious (ECO) \cite{ECO}, Bandit \cite{Bandit}, Square \cite{Square}, SignHunter \cite{SignHunter} and NES \cite{NES}. Since our method transforms the loss surface around $x$, it also applies to search-based attacks as well. Therefore, we also considered SimBA \cite{SimBA}. 

Some query-based attacks only consider hard label responses. Although our transformed loss surface (\ref{eq:defence}) does not affect these attacks, the additive noise from our complete attack in (\ref{eq:full_defence}) does \cite{RND}. Therefore, we include GeoDA (GeoD) \cite{GeoD}, SignOPT \cite{SignOPT} and HopSkipJump (HSJ) \cite{HopSkipJump} in our evaluation as well.

We used implementations found online\footnote{ \url{https://github.com/SCLBD/BlackboxBench}} and 
the hyperparamteres defined in \cite{RND,SignHunter}.
Following the experimental setup outlined in both those papers, we set the perturbation parameter $\epsilon=0.05$ for the $\ell_{\infty}$ norm across both datasets. For the $\ell_{2}$ norm, we adjusted $\epsilon$ to 1.0 and 5.0 for the CIFAR-10 and ImageNet datasets respectively. Following past works, all attacks were limited to 10,000 queries \cite{RND,SignHunter}. In this work, we only consider untargeted attacks. The results of the $\ell_{2}$ attacks can be found in the supplementary material.

\noindent\textbf{Baseline Defences.} As a baseline comparison to our method, we evaluated 6 preprocessors: JPEG Compression (JPEG) \cite{JpegCompression}, Uniform Noise (RND) \cite{RND}, Gaussian Noise (SND) \cite{GaussianNoiseSND}, Feature Squeezing (FS), Average Smoothing (AS) and Bit Squeezing (BS) \cite{FeatureSqueezing}. We note that RND and SND are considered state-of-the-art defences against query-based black box attacks. Although it is not a preprocessor, we also compare our performance to adversarial retraining (AT) \cite{AT} to provide a more complete view of our contribution.

We used the implementations found in the AdverTorch library\footnote{\url{https://github.com/BorealisAI/advertorch}} and in Adversarial Robustness Toolbox library.\footnote{\url{https://github.com/Trusted-AI/adversarial-robustness-toolbox}} Based on the experiments conducted by \cite{RND}, we utilized pretrained models from the MadryLab robustness library\footnote{\url{https://github.com/MadryLab/robustness}} for the AT models. Each model was pretrained on its respective dataset using a perturbation distance of $\epsilon=8/255$ for PGD.

\noindent\textbf{Metrics.} When defending against adversarial examples, it is important to consider (1) how many attacks are mitigated by the defence and (2) how much the model's performance suffers on clean data when the defence is in place. For the latter aspect, we measure the clean-task performance by calculating the model's accuracy (ACC) with and without the defence on clean data. For attack mitigation, we measure the attack failure rate (AFR) which is defined as the ratio of samples whose label did not change after the attack. To avoid bias, we only calculate the AFR on samples which the model can classify correctly before the attack.

\noindent\textbf{Experiments.}
We perform several experiments to evaluate the proposed defence. First we explore the impact of our defence's hyper parameters (learning rate $\alpha$ and the number of iterations $k$) on attack mitigation and clean data. Using these resutts, we select the best parameters and evaluate our defence against the black box attacks, and compare its performance to the baselines. Finally, we consider the robustness of the defense against an adaptive adversary which employs several different strategies. Additional experiments with different settings, such as the impact of using different epsilon and norms, can be found in the supplementary material.

Based on the hyperparameter experiment, we selected the most suitable hyperparameters for our defence and used them in all of our experiments. Unless otherwise noted, the hyperparameters were $k=10$ for the number of iterations and $\alpha=0.03$ and  $\alpha=0.1$ as the step size for CIFAR-10 and ImageNet respectively.

\section{Results}

\subsection{Defence Performance}\label{subsec:defensePerf}
In Table \ref{tab:baseline} we present the the effectiveness of our method in countering gradient and search-based attacks, as well as its performance against decision-based attacks relative to other defenses.

\begin{table}[t]
\centering
\caption{The performance of preprocessor defences when attacked by query-based black box attacks. The table shows the attack failure rate (AFR) for each defence, and how the model's accuracy is harmed by defence without any attacks (clean task performance). The best scores are in bold and second best scores are underlined.}
\label{tab:baseline}
\resizebox{\textwidth}{!}{%
\begin{tabular}{cc|c|cccccccccc|}
\rowcolor[HTML]{FFFFFF} 
                                                              & \textbf{Defence}           & \textit{Clean ACC} & \textit{NES}         & \textit{ZS}    & \textit{Bandit} & \textit{SignHunter} & \textit{ECO}   & \textit{SimBA}                                    & \textit{Sqaure} & \textit{SignOPT} & \textit{HSJ}         & \textit{GeoD}        \\ \hline
\rowcolor[HTML]{FFFFFF} 
\cellcolor[HTML]{FFFFFF}                                      & \textit{No Defence}      & 92.1\%             & 0                    & 0              & 0               & 0                   & 0              & 0                                                 & 0               & 0.024            & 0.152                & 0.13                 \\
\rowcolor[HTML]{FFFFFF} 
\cellcolor[HTML]{FFFFFF}                                      & \textit{AS}         & 42\%               & 0                    & 0.019          & 0               & 0                   & 0              & 0.073                                             & 0               & 0.163            & 0.334                & 0.218                \\
\rowcolor[HTML]{FFFFFF} 
\cellcolor[HTML]{FFFFFF}                                      & \textit{BS}         & 83\%               & 0.001                & 0              & 0.301           & 0.001               & 0              & 0.049                                             & 0.001           & 0.566            & 0.385                & 0.507                \\
\rowcolor[HTML]{FFFFFF} 
\cellcolor[HTML]{FFFFFF}                                      & \textit{JPEG}       & 79.4\%             & 0.0002               & 0              & 0.445           & 0.001               & 0.008          & 0.77                                              & 0               & 0.713            & \textbf{0.554}                & 0.631                \\
\rowcolor[HTML]{FFFFFF} 
\cellcolor[HTML]{FFFFFF}                                      & \textit{FS}         & 92\%               & 0                    & 0.002          & 0.475           & 0                   & 0              & 0.048                                             & 0               & 0.838            & {\ul 0.502}                & 0.552                \\
\rowcolor[HTML]{FFFFFF} 
\cellcolor[HTML]{FFFFFF}                                      & \textit{SND}   & 91.2\%               & 0.0008               & 0.006          & 0.410           & 0.187               & 0.536          & {\ul 0.843}                                             & 0.113           & \textbf{0.873}   & 0.484                & {\ul 0.653}                \\
\rowcolor[HTML]{FFFFFF} 
\cellcolor[HTML]{FFFFFF}                                      & \textit{RND}        & 87.3\%             & 0.011                & {\ul 0.039}          & 0.317           & 0.274               & 0.603          & 0.597                                             & 0.166           & {\ul 0.852}      & 0.439                & 0.590                \\
\rowcolor[HTML]{EFEFEF} 
\cellcolor[HTML]{FFFFFF}                                      & \textit{Ours k =1}  & 91.2\%             & {\ul 0.753}                & \textbf{0.730}          & \textbf{0.494}  & {\ul 0.620}         & {\ul 0.809}    & 0.833                                       & {\ul 0.464}     & \textbf{0.873}   & 0.483                & 0.652                \\
\rowcolor[HTML]{EFEFEF} 
\multirow{-9}{*}{\cellcolor[HTML]{FFFFFF} \rotatebox{90}{CIFAR-10 (Resnet20)}}  & \textit{Ours k =10} & \textbf{91.2\%}    & \textbf{0.764}       & \textbf{0.730}    & {\ul 0.491}     & \textbf{0.650}      & \textbf{0.810} & \textbf{0.861}                                    & \textbf{0.474}  & 0.848            & 0.485          & \textbf{0.662}          \\ \hline
\rowcolor[HTML]{FFFFFF} 
\cellcolor[HTML]{FFFFFF}                                      & \textit{No Defence}      & 75\%               & 0.07                 & 0.125          & 0.01            & 0                   & 0              & 0.237                                             & 0               & 0.162            & 0.6                  & 0.112                \\
\rowcolor[HTML]{FFFFFF} 
\cellcolor[HTML]{FFFFFF}                                      & \textit{AS}         & 68.8\%             & 0.069                & 0.477          & 0               & 0.002               & 0              & 0.43                                              & 0               & 0.834            & 0.68                 & 0.057                \\
\rowcolor[HTML]{FFFFFF} 
\cellcolor[HTML]{FFFFFF}                                      & \textit{BS}         & 66.2\%             & 0.132                & 0.405          & 0.352           & 0.002               & 0.003          & 0.176                                             & 0               & 0.903            & 0.75                 & 0.646                \\
\rowcolor[HTML]{FFFFFF} 
\cellcolor[HTML]{FFFFFF}                                      & \textit{JPEG}       & 73.8\%               & 0.139                & 0.5            & 0.501           & 0.004               & 0              & 0.215                                             & 0               & 0.925            & {\ul 0.788}          & 0.573                \\
\rowcolor[HTML]{FFFFFF} 
\cellcolor[HTML]{FFFFFF}                                      & \textit{FS}         & 74.6\%             & 0.075                & 0.485          & 0               & 0.004               & 0              & \multicolumn{1}{l}{\cellcolor[HTML]{FFFFFF}0.225} & 0               & 0.926            & 0.772                & 0.499                \\
\rowcolor[HTML]{FFFFFF} 
\cellcolor[HTML]{FFFFFF}                                      & \textit{SND}   & 74.3\%               & 0.075                & 0.691          & {\ul 0.531}     & 0.39                & 0.383          & {\ul 0.87}                                              & 0.131           & 0.952            & 0.756                & {\ul 0.68}           \\
\rowcolor[HTML]{FFFFFF} 
\cellcolor[HTML]{FFFFFF}                                      & \textit{RND}        & 72.8\%               & 0.11                 & 0.535          & 0.493           & 0.447               & 0.577          & \multicolumn{1}{l}{\cellcolor[HTML]{FFFFFF}0.797} & 0.241           & {\ul 0.956}      & 0.756                & 0.667                \\
\rowcolor[HTML]{EFEFEF} 
\cellcolor[HTML]{FFFFFF}                                      & \textit{Ours k =1}  & \textbf{74.3\%}    & {\ul 0.56}           & {\ul 0.752}    & \textbf{0.594}  & {\ul 0.706}         & {\ul 0.763}    & \cellcolor[HTML]{F1F1F1}\textbf{0.873}               & {\ul 0.475}     & 0.955            & 0.746                & 0.677                \\
\rowcolor[HTML]{EFEFEF} 
\multirow{-9}{*}{\cellcolor[HTML]{FFFFFF}\rotatebox{90}{ImageNet (Resnet50)}} & \textit{Ours k =10} & \textbf{74.3\%}    & {\textbf{0.658}} & \textbf{0.851} & \textbf{0.594}  & \textbf{0.761}      & \textbf{0.8}   & \cellcolor[HTML]{F1F1F1}{\textbf{0.873}}      & \textbf{0.519}  & \textbf{0.962}   & {\ul \textbf{0.794}} & {\ul \textbf{0.696}} \\ \hline
\end{tabular}%
}
\end{table}

Looking at the gradient/search-based attacks (NES, ZS, Bandit, SignHunter, ECo, SimBA and Square) our defense significantly outperforms all six baseline defences (rows) achieving an AFR as high as 0.86 and 0.87 on CIFAR-10 and ImageNet respectively. Our defense achieves an impressive average AFR of 0.68 and 0.72 on the respective datasets, across the evaluated attacks. Our method starkly outperforms the nearest competitors: SND and RND. SND has an average AFR of 0.3 and 0.46 on the datasets, while RND achieves an AFR of 0.29 and 0.46 on them respectively. When considering decision-based attacks (signOPT, HSJ and GeoD), our defense demonstrates performance on par with the state-of-the-art defenses RND and SND. As discussed in Section \ref{sec:meth}, decision-based attacks are not susceptible to the countermeasures involving optimization of the loss surface due to their reliance on the decision boundary rather than the gradient information. However, the addition of noise in our defense strategy still proves effective against these types of attacks, allowing our method to match the top-performing defenses in these scenarios.

The results also show that deployment of our defense does not noticeably affect the model's clean task performance, with a mere 0.7\% accuracy drop from the `no defense scenario'. Compared to other defences which suffer a 10-30\% performance reduction when there is no attack present. SND, while nearly matching our defense in terms of preserving clean task performance, falls short in achieving comparable robustness against these attacks.


Further extending our analysis, we compared our defense's performance with that of adversarial retraining (AT) \cite{AT}, a well-known and resource-intensive method that enhances model robustness by training on adversarial examples. We took the same model used in the evaluation of RND in \cite{RND}. The results of this comparison are presented in Table \ref{tab:AT}. Our defense outperforms AT in mitigating 6 out of 10 attacks while maintaining similar performance for the other 4 attacks. Crucially, \textbf{our method exhibits superior clean task performance compared to AT and does not require retraining a model}. This advantage is rooted in the inherent trade-off of AT, which sacrifices a degree of generalization to legitimate data to bolster defenses against adversarial examples \cite{madry2017towards}. Our defense, by design, aims to preserve the predicted class of inputs without altering the model's parameters, ensuring that the accuracy on clean tasks remains largely unaffected. This highlights our defense's unique capability to provide robust protection against adversarial threats while maintaining optimal performance on legitimate inputs.


\begin{table}[t]
\centering
\caption{The performance of our method (a preprocessor) in comparison with adversarial retraining (a model trainer). Bold indicates the best value.}
\label{tab:AT}
\resizebox{\textwidth}{!}{%
\begin{tabular}{cc|c|cccccccccc|}
& \textbf{Defence}                                   & \textit{Clean ACC}                      & \textit{NES}                           & \textit{ZS}                            & \textit{Bandit}                        & \textit{SignHunter}                    & \textit{ECO}                           & \textit{SimBA}                         & \textit{Sqaure}                        & \textit{SignOPT}                       & \textit{HSJ}                  & \textit{GeoD}                 \\ \hline
& \textit{No Defence}                              & 92.1\%                                  & 0                                      & 0                                      & 0                                      & 0                                      & 0                                      & 0                                      & 0                                      & 0.024                                  & 0.152                         & 0.13                          \\
& \textit{AT}                                 & 85\%                                    & 0.762                                  & \textbf{0.822}                         & \textbf{0.607}                         & 0.381                                  & 0.377                                  & 0.414                                  & 0.450                                  & 0.835                                  & \textbf{0.922}                & \textbf{0.734}                \\
\multirow{-3}{*}{\rotatebox{90}{CIFAR}}  & \cellcolor[HTML]{EFEFEF}\textit{Ours k =10} & \cellcolor[HTML]{EFEFEF}\textbf{91.2\%} & \cellcolor[HTML]{EFEFEF}\textbf{0.764} & \cellcolor[HTML]{EFEFEF}0.730          & \cellcolor[HTML]{EFEFEF}0.594          & \cellcolor[HTML]{EFEFEF}\textbf{0.650} & \cellcolor[HTML]{EFEFEF}\textbf{0.810} & \cellcolor[HTML]{EFEFEF}\textbf{0.861} & \cellcolor[HTML]{EFEFEF}\textbf{0.474} & \cellcolor[HTML]{EFEFEF}\textbf{0.848} & \cellcolor[HTML]{EFEFEF}0.485 & \cellcolor[HTML]{EFEFEF}0.662 \\ \hline
& \textit{No Defence}                              & 75\%                                    & 0.07                                   & 0.125                                  & 0.01                                   & 0                                      & 0                                      & 0.2375                                 & 0                                      & 0.162                                  & 0.6                           & 0.112                         \\
& \textit{AT}                                 & 47.7\%                                  & \textbf{0.706}                         & 0.771                                  & 0.465                                  & 0.433                                  & 0.312                                  & \textbf{0.949}                         & 0.392                                  & 0.924                                  & \textbf{0.926}                & \textbf{0.811}                \\
\multirow{-3}{*}{\rotatebox{90}{ImageN}} & \cellcolor[HTML]{EFEFEF}\textit{Ours k =10} & \cellcolor[HTML]{EFEFEF}\textbf{74.3\%} & \cellcolor[HTML]{EFEFEF}0.658          & \cellcolor[HTML]{EFEFEF}\textbf{0.851} & \cellcolor[HTML]{EFEFEF}\textbf{0.594} & \cellcolor[HTML]{EFEFEF}\textbf{0.761} & \cellcolor[HTML]{EFEFEF}\textbf{0.8}   & \cellcolor[HTML]{EFEFEF}0.873          & \cellcolor[HTML]{EFEFEF}\textbf{0.519} & \cellcolor[HTML]{EFEFEF}\textbf{0.962} & \cellcolor[HTML]{EFEFEF}0.794 & \cellcolor[HTML]{EFEFEF}0.696 \\ \hline
\end{tabular}%
}
\end{table}

\subsection{Hyper Parameters}
To optimize the effectiveness of the proposed defense against query-based black box attacks, we undertook a comprehensive evaluation focusing on varying step sizes ($\alpha$) the of iterations ($k$) for both the CIFAR-10 and ImageNet datasets.

\begin{figure}[t]
    \centering
    \includegraphics[width=\textwidth]{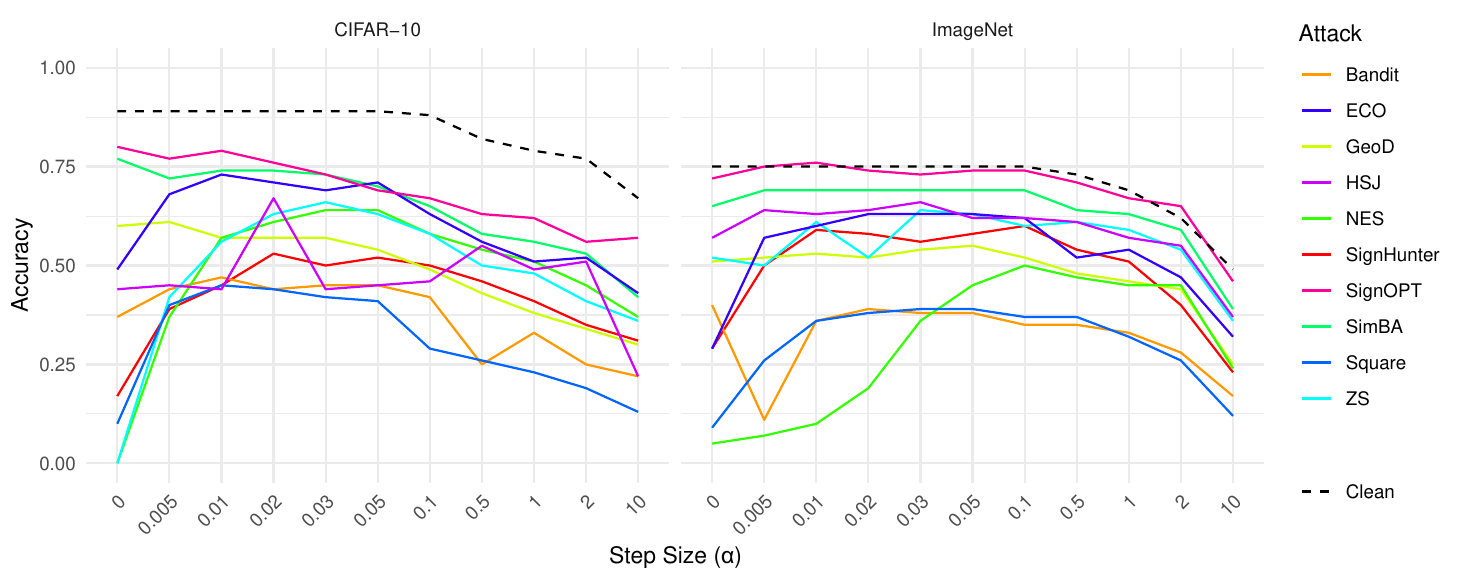}
    \caption{The affect of our defence's hyperparameter step-size ($\alpha$) on both clean data and adversarial examples.}
    \label{fig:step_size}
\end{figure}

First, we varied $\alpha$ with a fixed $k=10$. Figure \ref{fig:step_size} shows that using a small $\alpha$ (around 0.1) provides excellent results on both clean and malicious data. In this range, the clean task performance is barely affected too. However, increasing $\alpha$ beyond 0.1 causes the optimization to overshooting the target class $\hat{y}$, harming overall performance. To evaluate the impact of the $k$, we set $\alpha=0.1$ and evaluated $k$ ranging from 1 to 50. We found that $k$ does not significantly influence the defense's performance. This suggests that convergence is fast and that only one or two optimization steps are sufficient to significantly altering the local loss surface and disrupt the attack. This finding also highlights the practicality and scalability of counter-samples in real world settings. 

At $\alpha=0$, our defense essentially aligns with the SND approach, relying solely on adding Gaussian noise without further optimization. The transition from $\alpha=0$ to $\alpha=0.1$ marks a significant leap in model robustness for all gradient based attacks as discussed earlier. 

\subsection{Adaptive Adversary}
Tramer et al. \cite{AdaptiveAttacker} raises the need for evaluating defences in the presence of adaptive adversaries. 
An adaptive adversary refers to an attacker who specifically tailors their strategies to overcome a system's defenses by exploiting knowledge about the deployed defense mechanisms. One form of adaptive adversary is to simply attack a model \textit{with} the defence in place.  Our evaluation in Section \ref{subsec:defensePerf} already captures this scenario and shows that our model is robust to this form of adaptive adversary.
In this section we consider two other strategies customized to defeat our defence.

\textbf{Strategy 1 (Averaging Queries).}
The first strategy is to counter the random noise we add in (\ref{eq:full_defence}). The attack is to send each query $M$ times and average the result. This strategy was proposed by Athalye et al. in \cite{athalye2018synthesizing} and shown to be effective against defences that utilize random noise \cite{RND}. There are two reasons why this is a reasonable strategy against counter-samples: (1) we always add random noise $z$ to $x$ before performing optimization and (2) the result $x^*$ will be affected by the starting point $x+z$. Therefore, by averaging repeated queries it may be possible to mitigate the benefit of the added noise. 
To evaluate our defence against this adaptive adversary, we attacked our defence with $M=\{1,5,10\}$ based on \cite{RND}. To ensure a fair evaluation, we let the attacks perform $10,000*M$ queries in total. While increasing the query factor beyond $M=10$ may help mitigate the defences further, the strategy becomes less feasible due to the importance of query efficiency in maintaining stealth in black box attacks. Query counts from our experiments can be found in the supplementary material.

In Table \ref{tab:adapt1} we present the results from this experiment. The table underscores our method's robustness to such adaptive tactics. While state-of-the-art noise-based preprocessors like RND and SND see their AFR significantly drop by as much as 0.26 and 0.45 respectively under these strategies, our defense's AFR reduction is capped at about 0.1. Unlike RND and SND, which exhibit a clear trend of degradation as $M$ increases, our defense's performance remains largely stable, with AFR fluctuating and, in some cases, even improving. This suggests that while averaging may counteract our noising step, it fails to undermine the optimization step, potentially setting a lower bound for our defense's efficacy. 




\begin{table}[t]
\centering
\caption{The performance of our defence when under attack by an adaptive adversary using Strategy 1: Each query is repeated $M$ times and then averaged. RND and SND are other preprocessors that are noise based. Results are reported in AFR.}
\label{tab:adapt1}
\resizebox{0.85\textwidth}{!}{%
\begin{tabular}{ccccccccccc}
                                                & \multicolumn{1}{c|}{}                          & \multicolumn{3}{c|}{\textit{RND}}                                         & \multicolumn{3}{c|}{\cellcolor[HTML]{FFFFFF}\textit{SND}}                 & \multicolumn{3}{c}{\cellcolor[HTML]{EFEFEF}\textit{Ours $k = 10$}}                                 \\ \hline
\multicolumn{1}{c|}{}                           & \multicolumn{1}{c|}{Attack} & M = 1                & M = 5                & \multicolumn{1}{c|}{M = 10} & M = 1                & M = 5                & \multicolumn{1}{c|}{M = 10} & \cellcolor[HTML]{EFEFEF}M = 1 & \cellcolor[HTML]{EFEFEF}M = 5  & \cellcolor[HTML]{EFEFEF}M = 10 \\ \hline
\multicolumn{1}{c|}{}                           & \multicolumn{1}{c|}{\textit{Bandit}}           & 0.329                & 0.298                    & \multicolumn{1}{c|}{0.252}  & 0.465                & 0.386                & \multicolumn{1}{c|}{0.352}  & \cellcolor[HTML]{EFEFEF}0.491 & \cellcolor[HTML]{EFEFEF}0.471  & \cellcolor[HTML]{EFEFEF}0.507  \\
\multicolumn{1}{c|}{}                           & \multicolumn{1}{c|}{\textit{NES}}              & 0.011                & 0                    & \multicolumn{1}{c|}{0}      & 0.0001               & 0                    & \multicolumn{1}{c|}{0}      & \cellcolor[HTML]{EFEFEF}0.764 & \cellcolor[HTML]{EFEFEF}0.715  & \cellcolor[HTML]{EFEFEF}0.715  \\
\multicolumn{1}{c|}{}                           & \multicolumn{1}{c|}{\textit{ECO}}              & 0.634                & 0.517                & \multicolumn{1}{c|}{0.341}  & 0.554                & 0.168                & \multicolumn{1}{c|}{0.103}  & \cellcolor[HTML]{EFEFEF}0.81  & \cellcolor[HTML]{EFEFEF}0.797  & \cellcolor[HTML]{EFEFEF}0.825  \\
\multicolumn{1}{c|}{}                           & \multicolumn{1}{c|}{\textit{SignHunter}}       & 0.306                & 0.114                & \multicolumn{1}{c|}{0.0112} & 0.204                & 0.022                & \multicolumn{1}{c|}{0}      & \cellcolor[HTML]{EFEFEF}0.651 & \cellcolor[HTML]{EFEFEF}0.643  & \cellcolor[HTML]{EFEFEF}0.551  \\
\multicolumn{1}{c|}{}                           & \multicolumn{1}{c|}{\textit{SimBA}}            & 0.597                & 0.613               & \multicolumn{1}{c|}{0.586}  & 0.41                & 0.829                & \multicolumn{1}{c|}{0.818}  & \cellcolor[HTML]{EFEFEF}0.854 & \cellcolor[HTML]{EFEFEF}0.839  & \cellcolor[HTML]{EFEFEF}0.838  \\
\multicolumn{1}{c|}{}                           & \multicolumn{1}{c|}{\textit{Square}}           & 0.136                & 0.045                & \multicolumn{1}{c|}{0.022}  & 0.136                & 0                    & \multicolumn{1}{c|}{0}      & \cellcolor[HTML]{EFEFEF}0.474 & \cellcolor[HTML]{EFEFEF}0.505  & \cellcolor[HTML]{EFEFEF}0.489  \\
\multicolumn{1}{c|}{\multirow{-7}{*}{\rotatebox{90}{CIFAR-10}}}  & \multicolumn{1}{c|}{\textit{ZS}}               & 0.068                & 0                    & \multicolumn{1}{c|}{0}      & 0.005                & 0                    & \multicolumn{1}{c|}{0}      & \cellcolor[HTML]{EFEFEF}0.73  & \cellcolor[HTML]{EFEFEF}0.735  & \cellcolor[HTML]{EFEFEF}0.747  \\ \hline
\multicolumn{1}{c|}{}                           & \multicolumn{1}{c|}{\textit{Bandit}}           & 0.493                & 0.438                    & \multicolumn{1}{c|}{0.392}      & 0.531                 & 0.493                    & \multicolumn{1}{c|}{0.412}      & \cellcolor[HTML]{EFEFEF}0.594 & \cellcolor[HTML]{EFEFEF}0.603  & \cellcolor[HTML]{EFEFEF}0.606  \\
\multicolumn{1}{c|}{}                           & \multicolumn{1}{c|}{\textit{NES}}              & 0.09                 & 0.027                & \multicolumn{1}{c|}{0}      & 0.075                & 0.025                & \multicolumn{1}{c|}{0}      & \cellcolor[HTML]{EFEFEF}0.658 & \cellcolor[HTML]{EFEFEF}0.772  & \cellcolor[HTML]{EFEFEF}0.91   \\
\multicolumn{1}{c|}{}                           & \multicolumn{1}{c|}{\textit{ECO}}              & 0.602                & 0.354                & \multicolumn{1}{c|}{0.126}  & 0.43                 & 0.137                & \multicolumn{1}{c|}{0.081}  & \cellcolor[HTML]{EFEFEF}0.8   & \cellcolor[HTML]{EFEFEF}0.833  & \cellcolor[HTML]{EFEFEF}0.807  \\
\multicolumn{1}{c|}{}                           & \multicolumn{1}{c|}{\textit{SignHunter}}       & 0.532                & 0.27                 & \multicolumn{1}{c|}{0.162}  & 0.379                & 0.205                & \multicolumn{1}{c|}{0.153}  & \cellcolor[HTML]{EFEFEF}0.761 & \cellcolor[HTML]{EFEFEF}0.721  & \cellcolor[HTML]{EFEFEF}0.759  \\
\multicolumn{1}{c|}{}                           & \multicolumn{1}{c|}{\textit{SimBA}}            & 0.797                & 0.796                & \multicolumn{1}{c|}{0.796}  & 0.87                & 0.86                 & \multicolumn{1}{c|}{0.864}   & \cellcolor[HTML]{EFEFEF}0.873 & \cellcolor[HTML]{EFEFEF}0.85   & \cellcolor[HTML]{EFEFEF}0.862  \\
\multicolumn{1}{c|}{}                           & \multicolumn{1}{c|}{\textit{Square}}           & 0.227                & 0.064                & \multicolumn{1}{c|}{0.038}  & 0.101                & 0.025                & \multicolumn{1}{c|}{0}      & \cellcolor[HTML]{EFEFEF}0.519 & \cellcolor[HTML]{EFEFEF}0.544  & \cellcolor[HTML]{EFEFEF}0.544  \\
\multicolumn{1}{c|}{\multirow{-7}{*}{\rotatebox{90}{ImageNet}}} & \multicolumn{1}{c|}{\textit{ZS}}               & 0.556                & 0.297                & \multicolumn{1}{c|}{0.194}  & 0.253                & 0.128                & \multicolumn{1}{c|}{0.079}  & \cellcolor[HTML]{EFEFEF}0.851 & \cellcolor[HTML]{EFEFEF}0.8125 & \cellcolor[HTML]{EFEFEF}0.8375 \\ \hline
\multicolumn{1}{l}{}                            & \multicolumn{1}{l}{}                           & \multicolumn{1}{l}{} & \multicolumn{1}{l}{} & \multicolumn{1}{l}{}        & \multicolumn{1}{l}{} & \multicolumn{1}{l}{} & \multicolumn{1}{l}{}        & \multicolumn{1}{l}{}          & \multicolumn{1}{l}{}           & \multicolumn{1}{l}{}           \\
\multicolumn{1}{l}{}                            & \multicolumn{1}{l}{}                           & \multicolumn{1}{l}{} & \multicolumn{1}{l}{} & \multicolumn{1}{l}{}        & \multicolumn{1}{l}{} & \multicolumn{1}{l}{} & \multicolumn{1}{l}{}        & \multicolumn{1}{l}{}          & \multicolumn{1}{l}{}           & \multicolumn{1}{l}{}          
\end{tabular}%
}
\end{table}



\textbf{Strategy 2 (Increasing Step Size).}
The second strategy is to try and counter our optimization process in (\ref{eq:defence}). As outlined in Section \ref{SumGame}, we are essentially countering the attacker's optimization objective with an opposite objective. The attack is to increase the step size $\alpha$ of the attack algorithms to be larger than the step size used by our defence. The concept is that by increasing the attacker's step size, it might be feasible to alter $\hat{y}$ at an early stage, thereby disrupting the defender's optimization objective, despite the defender's ability to take multiple whitebox steps for each blackbox step executed by the attacker.

For this experiment we examined ZS, Bandit and NES as gradient-based attacks and SimBA as a search-based attack. We evaluated each attack with a step size increased by a factor of 1,  2 and 10. Table \ref{tab:adapt2} shows that overall, our method out performs the other baselines. For ImageNet, our defence remains robust to increased steps sizes \textbf{for all of the attacks} since the AFR increases when reaching a step size factor of 10. This is probably because ImageNet has a large resolution  making the number of optimization iterations more important than the step size in this context. For CIFAR-10, increasing the step size seems to have an effect on our defence when using ZS and NES. However, our method still outperform other baselines and is still not affected by the Bandit and SimBA attacks.

\begin{table}[t]
\centering
\caption{The performance of our defence when under attack by an adaptive adversary using Strategy 2: adversaries use much larger steps sizes than usual to counter our defence's fixed step size. The learning rates have been increased by a factor higher than their default values as listed in the second row. The best score is bold, the second best is underlined.}
\label{tab:adapt2}

\resizebox{\textwidth}{!}{%
\begin{tabular}{cccccccccccccccccccccccccc}
 
 &  &  &  &  &  &  &  & \multicolumn{1}{l|}{}                                      & \multicolumn{1}{l|}{}                                            & \multicolumn{3}{c}{ZS}                                                                                                                                     & \multicolumn{3}{c|}{Bandit}                                                                                                                                       & \multicolumn{3}{c|}{NES}                                                                                                                                          & \multicolumn{3}{c}{SimBA}                                                                                                                                         &  &  &  &  \\ \cline{9-22}
 &  &  &  &  &  &  &  & \multicolumn{1}{l|}{}                                      & \multicolumn{1}{c|}{Defence}                            & $\times 1$                                                         & $\times 2$                                     & \multicolumn{1}{c|}{$\times 10$}                                    & $\times 1$                                                        & $\times 2$                                     & \multicolumn{1}{c|}{$\times 10$}                                    & $\times 1$                                                         & $\times 2$                                     & \multicolumn{1}{c|}{$\times 10$}                                    & $\times 1$                                                         & $\times 2$                                     & \multicolumn{1}{c|}{$\times 10$}                                    &  &  &  &  \\ \cline{10-22}
 &  &  &  &  &  &  &  & \multicolumn{1}{l|}{}                                      & \multicolumn{1}{c|}{\textit{AS}}                                 & \multicolumn{1}{c}{0}                                      & 0.04                                   & \multicolumn{1}{c|}{0.146}                                  & \multicolumn{1}{c}{0}                                      & 0                                      & \multicolumn{1}{c|}{0}                                      & \multicolumn{1}{c}{0}                                      & 0.024                                  & \multicolumn{1}{c|}{0.219}                                  & \multicolumn{1}{c}{0.073}                                  & 0                                      & \multicolumn{1}{c|}{0}                                      &  &  &  &  \\
 &  &  &  &  &  &  &  & \multicolumn{1}{l|}{}                                      & \multicolumn{1}{c|}{\textit{BS}}                                 & \multicolumn{1}{c}{0.001}                                  & 0.06                                   & \multicolumn{1}{c|}{0.192}                                  & \multicolumn{1}{c}{0.301}                                  & 0.301                                  & \multicolumn{1}{c|}{0.409}                                  & \multicolumn{1}{c}{0.001}                                  & 0.036                                  & \multicolumn{1}{c|}{0.168}                                  & \multicolumn{1}{c}{0.049}                                  & 0                                      & \multicolumn{1}{c|}{0}                                      &  &  &  &  \\
 &  &  &  &  &  &  &  & \multicolumn{1}{l|}{}                                      & \multicolumn{1}{c|}{\textit{JPEG}}                               & \multicolumn{1}{c}{0.0002}                                 & 0                                      & \multicolumn{1}{c|}{0.162}                                  & \multicolumn{1}{c}{0.445}                                  & 0.35                                   & \multicolumn{1}{c|}{0.378}                                  & \multicolumn{1}{c}{0.0002}                                 & 0                                      & \multicolumn{1}{c|}{0.175}                                  & \multicolumn{1}{c}{0.77}                                   & 0.581                                  & \multicolumn{1}{c|}{0.108}                                  &  &  &  &  \\
 &  &  &  &  &  &  &  & \multicolumn{1}{l|}{}                                      & \multicolumn{1}{c|}{\textit{FS}}                                 & \multicolumn{1}{c}{0}                                      & 0.032                                  & \multicolumn{1}{c|}{0.184}                                  & {\ul 0.475}                                                & 0.321                                  & \multicolumn{1}{c|}{0.367}                                  & \multicolumn{1}{c}{0}                                      & 0                                      & \multicolumn{1}{c|}{0.163}                                  & \multicolumn{1}{c}{0.048}                                  & 0                                      & \multicolumn{1}{c|}{0}                                      &  &  &  &  \\
 &  &  &  &  &  &  &  & \multicolumn{1}{l|}{}                                      & \multicolumn{1}{c|}{\textit{SND}}                                & \multicolumn{1}{c}{0.0008}                                 & 0.08                                   & \multicolumn{1}{c|}{\textbf{0.229}}                         & \multicolumn{1}{c}{0.465}                                  & {\ul 0.505}                            & \multicolumn{1}{c|}{{\ul 0.378}}                            & 0.0001                                                     & 0.02                                   & \multicolumn{1}{c|}{0.215}                                  & \multicolumn{1}{c}{{\ul 0.843}}                            & {\ul 0.845}                            & \multicolumn{1}{c|}{{\ul 0.841}}                            &  &  &  &  \\
 &  &  &  &  &  &  &  & \multicolumn{1}{l|}{}                                      & \multicolumn{1}{c|}{\textit{RND}}                                & {\ul 0.068}                                                & {\ul 0.164}                            & \multicolumn{1}{c|}{0.202}                                  & \multicolumn{1}{c}{0.329}                                  & 0.364                                  & \multicolumn{1}{c|}{0.482}                                  & {\ul 0.011}                                                & {\ul 0.116}                            & \multicolumn{1}{c|}{{\ul 0.202}}                            & \multicolumn{1}{c}{0.597}                                  & 0.59                                   & \multicolumn{1}{c|}{0.61}                                   &  &  &  &  \\
 &  &  &  &  &  &  &  & \multicolumn{1}{l|}{\multirow{-7}{*}{\rotatebox{90}{CIFAR-10}}}  & \multicolumn{1}{c|}{\cellcolor[HTML]{EFEFEF}\textit{Ours $k =10$}} & \multicolumn{1}{c}{\cellcolor[HTML]{EFEFEF}\textbf{0.764}} & \cellcolor[HTML]{EFEFEF}\textbf{0.448} & \multicolumn{1}{c|}{\cellcolor[HTML]{EFEFEF}{\ul 0.206}}    & \multicolumn{1}{c}{\cellcolor[HTML]{EFEFEF}\textbf{0.491}} & \cellcolor[HTML]{EFEFEF}\textbf{0.54}  & \multicolumn{1}{c|}{\cellcolor[HTML]{EFEFEF}\textbf{0.517}} & \multicolumn{1}{c}{\cellcolor[HTML]{EFEFEF}\textbf{0.764}} & \cellcolor[HTML]{EFEFEF}\textbf{0.465} & \multicolumn{1}{c|}{\cellcolor[HTML]{EFEFEF}\textbf{0.204}} & \multicolumn{1}{c}{\cellcolor[HTML]{EFEFEF}\textbf{0.861}} & \cellcolor[HTML]{EFEFEF}\textbf{0.847} & \multicolumn{1}{c|}{\cellcolor[HTML]{EFEFEF}\textbf{0.845}} &  &  &  &  \\ \cline{9-22}
 &  &  &  &  &  &  &  & \multicolumn{1}{l|}{}                                      & \multicolumn{1}{c|}{\textit{AS}}                                 & \multicolumn{1}{c}{0.477}                                  & 0.236                                  & \multicolumn{1}{c|}{0.51}                                   & \multicolumn{1}{c}{\cellcolor[HTML]{FFFFFF}0}              & \cellcolor[HTML]{FFFFFF}0              & \multicolumn{1}{c|}{\cellcolor[HTML]{FFFFFF}0}              & \multicolumn{1}{c}{0.069}                                  & 0.097                                  & \multicolumn{1}{c|}{0.361}                                  & \multicolumn{1}{c}{\cellcolor[HTML]{FFFFFF}0.43}           & \cellcolor[HTML]{FFFFFF}0.25           & \multicolumn{1}{c|}{\cellcolor[HTML]{FFFFFF}0.43}           &  &  &  &  \\
 &  &  &  &  &  &  &  & \multicolumn{1}{l|}{}                                      & \multicolumn{1}{c|}{\textit{BS}}                                 & \multicolumn{1}{c}{0.405}                                  & 0.294                                  & \multicolumn{1}{c|}{0.705}                                  & \multicolumn{1}{c}{\cellcolor[HTML]{FFFFFF}0.352}          & \cellcolor[HTML]{FFFFFF}0.397          & \multicolumn{1}{c|}{\cellcolor[HTML]{FFFFFF}0.382}          & \multicolumn{1}{c}{0.132}                                  & 0.147                                  & \multicolumn{1}{c|}{0.441}                                  & \multicolumn{1}{c}{\cellcolor[HTML]{FFFFFF}0.176}          & \cellcolor[HTML]{FFFFFF}0.191          & \multicolumn{1}{c|}{\cellcolor[HTML]{FFFFFF}0.176}          &  &  &  &  \\
 &  &  &  &  &  &  &  & \multicolumn{1}{l|}{}                                      & \multicolumn{1}{c|}{\textit{JPEG}}                               & \multicolumn{1}{c}{0.5}                                    & 0.379                                  & \multicolumn{1}{c|}{{\ul 0.721}}                            & \multicolumn{1}{c}{\cellcolor[HTML]{FFFFFF}0.506}          & \cellcolor[HTML]{FFFFFF}0.492          & \multicolumn{1}{c|}{\cellcolor[HTML]{FFFFFF}{\ul 0.518}}    & \multicolumn{1}{c}{{\ul 0.139}}                            & {\ul 0.177}                            & \multicolumn{1}{c|}{0.518}                                  & \multicolumn{1}{c}{\cellcolor[HTML]{FFFFFF}0.215}          & \cellcolor[HTML]{FFFFFF}0.191          & \multicolumn{1}{c|}{\cellcolor[HTML]{FFFFFF}0.215}          &  &  &  &  \\
 &  &  &  &  &  &  &  & \multicolumn{1}{l|}{}                                      & \multicolumn{1}{c|}{\textit{FS}}                                 & \multicolumn{1}{c}{0.485}                                  & 0.225                                  & \multicolumn{1}{c|}{0.6}                                    & \multicolumn{1}{c}{\cellcolor[HTML]{FFFFFF}0}              & \cellcolor[HTML]{FFFFFF}0              & \multicolumn{1}{c|}{\cellcolor[HTML]{FFFFFF}0}              & \multicolumn{1}{c}{0.075}                                  & 0.125                                  & \multicolumn{1}{c|}{0.48}                                   & \multicolumn{1}{c}{\cellcolor[HTML]{FFFFFF}0.2375}         & \cellcolor[HTML]{FFFFFF}0.225          & \multicolumn{1}{c|}{\cellcolor[HTML]{FFFFFF}0.23}           &  &  &  &  \\
 &  &  &  &  &  &  &  & \multicolumn{1}{l|}{}                                      & \multicolumn{1}{c|}{\textit{SND}}                                & \multicolumn{1}{c}{{\ul 0.691}}                            & 0.518                                  & \multicolumn{1}{c|}{0.696}                                  & \multicolumn{1}{c}{\cellcolor[HTML]{FFFFFF}{\ul 0.531}}    & \cellcolor[HTML]{FFFFFF}{\ul 0.481}    & \multicolumn{1}{c|}{\cellcolor[HTML]{FFFFFF}0.479}          & \multicolumn{1}{c}{0.075}                                  & 0.113                                  & \multicolumn{1}{c|}{0.531}                                  & \cellcolor[HTML]{FFFFFF}{\ul 0.87}                         & \cellcolor[HTML]{FFFFFF}{\ul 0.873}    & \multicolumn{1}{c|}{\cellcolor[HTML]{FFFFFF}{\ul 0.85}}     &  &  &  &  \\
 &  &  &  &  &  &  &  & \multicolumn{1}{l|}{}                                      & \multicolumn{1}{c|}{\textit{RND}}                                & \multicolumn{1}{c}{0.535}                                  & {\ul 0.62}                             & \multicolumn{1}{c|}{0.779}                                  & \multicolumn{1}{c}{\cellcolor[HTML]{FFFFFF}0.493}          & \cellcolor[HTML]{FFFFFF}0.455          & \multicolumn{1}{c|}{\cellcolor[HTML]{FFFFFF}0.43}           & 0.038                                                      & 0.129                                  & \multicolumn{1}{c|}{{\ul 0.545}}                            & \cellcolor[HTML]{FFFFFF}0.797                              & \cellcolor[HTML]{FFFFFF}0.772          & \multicolumn{1}{c|}{\cellcolor[HTML]{FFFFFF}0.784}          &  &  &  &  \\
 &  &  &  &  &  &  &  & \multicolumn{1}{l|}{\multirow{-7}{*}{\rotatebox{90}{ImageNet}}} & \multicolumn{1}{c|}{\cellcolor[HTML]{EFEFEF}\textit{Ours $k =10$}} & \cellcolor[HTML]{EFEFEF}\textbf{0.851}                     & \cellcolor[HTML]{EFEFEF}\textbf{0.772} & \multicolumn{1}{c|}{\cellcolor[HTML]{EFEFEF}\textbf{0.784}} & \multicolumn{1}{c}{\cellcolor[HTML]{EFEFEF}\textbf{0.594}} & \cellcolor[HTML]{EFEFEF}\textbf{0.55} & \multicolumn{1}{c|}{\cellcolor[HTML]{EFEFEF}\textbf{0.57}}  & \multicolumn{1}{c}{\cellcolor[HTML]{EFEFEF}\textbf{0.658}} & \cellcolor[HTML]{EFEFEF}\textbf{0.383} & \multicolumn{1}{c|}{\cellcolor[HTML]{EFEFEF}\textbf{0.759}} & \cellcolor[HTML]{EFEFEF}\textbf{0.873}                     & \cellcolor[HTML]{EFEFEF}\textbf{0.86}  & \multicolumn{1}{c|}{\cellcolor[HTML]{EFEFEF}\textbf{0.89}}  &  &  &  &  \\ 
\end{tabular}%
}
\end{table}

In summary, we affirm the critical importance of designing defense mechanisms that are not only effective against known attack vectors but also adaptable and resilient in the face of adversarial strategies. Our findings demonstrate that, although there may be scenarios where adaptive adversaries can challenge the proposed defense, it consistently exhibits robustness against such tactics and surpasses the resilience of other preprocessor defenses. This resilience underscores our defense's capacity to adapt and protect machine learning models in a landscape where adversaries continuously evolve their attack methodologies. Ultimately, our approach contributes significantly to the ongoing effort to fortify machine learning systems against the sophisticated and adaptive threats posed by modern adversaries. 

In this paper, we presented one possible implementation of counter-samples, showcasing its efficacy and adaptability against a range of adversarial attacks. We envision this work as a foundational step that can be further refined and expanded by other researchers to develop even more robust defenses in the rapidly evolving field of adversarial machine learning.




\section{Conclusion}
In this paper, we introduced a novel defense mechanism against query-based black box attacks, centered around the innovative concept of counter-samples. This approach diverges from traditional preprocessing methods by directly challenging the attack process, thereby creating a favorable asymmetry for the defender. Our method is distinguished by its effectiveness in misleading the attacker's search for adversarial examples, while simultaneously preserving the accuracy of the model on legitimate inputs across a broad spectrum of attack types. While being effective against gradient-based and search based attacks, the method is also effective against decision-based attacks as well.

Through rigorous evaluation against state-of-the-art black box attacks on CIFAR-10 and ImageNet datasets, we demonstrated that our defense provide state-of-the-art performance \textit{without} harming the model performance on benign samples. Moreover, we showed how the method is robustness against strong adversaries compared to existing defences. Additionally, the strategy's stateless nature ensures its scalability and ease of integration into existing systems without the complexities of tracking user queries or significant computational burdens.

This paper lays the groundwork for future research to explore and refine dynamic, stateless counter-sample strategies, opening new avenues for enhancing the security and integrity of machine learning models in the face of evolving adversarial challenges.

\bibliographystyle{splncs}
\bibliography{egbib}
\end{document}